%% file: main.tex
\documentclass[sigconf,authorversion,nonacm]{acmart}

\pdfoutput=1 

\input{section/preamble/preamble}
\input{section/preamble/preamblelistingsconf}
\input{section/preamble/acronyms}
\input{section/preamble/macros}

\begin{document}

\title{Don't CWEAT It: Toward {CWE} {A}nalysis {T}echniques in Early Stages of Hardware Design
}

\author{Baleegh Ahmad}
\orcid{0000-0001-6854-3966}
\affiliation{%
  \institution{New York University}
  \country{}
}

\author{Wei-Kai Liu}
\orcid{1234-5678-9012}
\affiliation{%
  \institution{Duke University}
    \country{}
}

\author{Luca Collini}
\orcid{1234-5678-9012}
\affiliation{%
  \institution{New York University}
    \country{}
}

\author{Hammond Pearce}
\orcid{0000-0002-3488-7004}
\affiliation{%
  \institution{New York University}
    \country{}
}

\author{Jason M. Fung}
\affiliation{%
  \institution{Intel Corporation}
    \country{}
}
\author{Jonathan Valamehr}
\affiliation{%
  \institution{Intel Corporation}
    \country{}
}
\author{Mohammad Bidmeshki}
\affiliation{%
  \institution{Intel Corporation}
    \country{}
}
\author{Piotr Sapiecha}
\orcid{0000-0001-8757-4848}
\affiliation{%
  \institution{Intel Corporation}
    \country{}
}
\author{Steve Brown}
\affiliation{%
  \institution{Intel Corporation}
    \country{}
}

\author{Krishnendu Chakrabarty}
\orcid{1234-5678-9012}
\affiliation{%
  \institution{Duke University}
    \country{}
}

\author{Ramesh Karri}
\orcid{1234-5678-9012}
\affiliation{%
  \institution{New York University}
    \country{}
}

\author{Benjamin Tan}
\orcid{0000-0002-7642-3638}
\affiliation{%
  \institution{University of Calgary}
    \country{}
}

\begin{abstract}
To help prevent hardware security vulnerabilities from propagating to later design stages where fixes are costly, it is crucial to identify security concerns as early as possible, such as in RTL designs. 
In this work, we investigate the practical implications and feasibility of producing a set of security-specific scanners that operate on Verilog source files. 
The scanners indicate parts of code that might contain one of a set of MITRE's common weakness enumerations (CWEs). 
We explore the CWE database to characterize the scope and attributes of the CWEs and identify those that are amenable to static analysis. We prototype scanners and evaluate them on 11 open source designs -- 4 system-on-chips (SoC) and 7 processor cores -- and explore the nature of identified weaknesses. Our analysis reported 53 potential weaknesses in the OpenPiton SoC used in Hack@DAC-21, 11 of which we confirmed as security concerns.

\end{abstract}






\maketitle

\thispagestyle{plain}
\pagestyle{plain}


\input{section/01Introduction}

\input{section/01x_Background}

\input{section/02StaticAnalysisCWEs}

\input{section/03Algorithms}
\input{section/04Results}

\input{section/05SecurityAssesment}

\input{section/06Discussion}

\input{section/07Conclusion}


\clearpage
\bibliographystyle{ACM-Reference-Format}
\bibliography{lit/benhamram.bib}


\end{document}

%% file: section/preamble/preamble.tex
\usepackage[utf8]{inputenc} 
\usepackage[most]{tcolorbox} 

\usepackage{multirow} 
\usepackage[shortlabels]{enumitem}  

\usepackage{subcaption} 
\usepackage{pgfplots} 
\usepackage{pgfplotstable} 
\usepgfplotslibrary{statistics} 

\usepackage{comment} 
\usepackage{listings}  
\definecolor{vgreen}{RGB}{104,180,104}
\definecolor{vblue}{RGB}{49,49,255}
\definecolor{vorange}{RGB}{255,143,102}

\lstdefinestyle{verilog-style}
{
    language=Verilog,
    basicstyle=\small\ttfamily,
    keywordstyle=\color{vblue},
    identifierstyle=\color{black},
    commentstyle=\color{vgreen},
    numbers=left,
    numberstyle=\tiny\color{black},
    numbersep=10pt,
    tabsize=8,
    moredelim=*[s][\colorIndex]{[}{]},
    literate=*{:}{:}1
}

\makeatletter
\newcommand*\@lbracket{[}
\newcommand*\@rbracket{]}
\newcommand*\@colon{:}
\newcommand*\colorIndex{%
    \edef\@temp{\the\lst@token}%
    \ifx\@temp\@lbracket \color{black}%
    \else\ifx\@temp\@rbracket \color{black}%
    \else\ifx\@temp\@colon \color{black}%
    \else \color{vorange}%
    \fi\fi\fi
}
\makeatother

\usepackage{trace} 

\usepackage{circledsteps} 

\usetikzlibrary{patterns}

\usepackage{acronym} 
\input{section/preamble/acronyms}

\usepackage{colortbl} 

\usepackage{float} 
\usepackage{textcomp} 

\usepackage{algorithm} 
\usepackage{algpseudocode}  

\newcommand\CONDITION[2]%
  {\begin{tabular}[t]{@{}l@{}l@{}}
     #1&#2
   \end{tabular}%
  }

\algdef{SE}[IF]{If}{EndIf}[1]%
  {\algorithmicif\ \CONDITION{#1}{\ \algorithmicthen}}%
  {\algorithmicend\ \algorithmicif}%
\usepackage{url}

\usepackage{pifont} 
\usepackage{framed} 
\usepackage{soul} 
\usepackage{multirow} 
\newcolumntype{L}[1]{>{\raggedright\let\newline\\\arraybackslash\hspace{0pt}}m{#1}}
\newcolumntype{C}[1]{>{\centering\let\newline\\\arraybackslash\hspace{0pt}}m{#1}}
\newcolumntype{R}[1]{>{\raggedleft\let\newline\\\arraybackslash\hspace{0pt}}m{#1}}
\newcolumntype{H}{>{\collectcell\lstinline}l<{\endcollectcell}}
\usepackage{url} 
\newcommand{\todoblock}[1]{\noindent\definecolor{shadecolor}{RGB}{252, 217, 217}\colorbox{shadecolor}{\parbox{\columnwidth}{{#1}}}}
\newcommand{\todoblockblue}[1]{\noindent\definecolor{shadecolor}{RGB}{214, 254, 255}\colorbox{shadecolor}{\parbox{\columnwidth}{{#1}}}}

\newcommand{\rt}[1]{{\color{red}{#1}}}
\newcommand{\redtxt}[1]{{\color{red}{#1}}}
\newcommand{\greentxt}[1]{{\color{olive}{#1}}}

\newcommand{\cmark}{\ding{51}}%
\newcommand{\xmark}{\ding{55}}%

\newcommand{\unbf}[1]{\underline{\textbf{#1}}}

\newcommand{\sol}{\texttt{CWEAT}}


%% file: section/preamble/acronyms.tex
\acrodef{CPS}{Cyber-Physical System}
\acrodef{IoT}{Internet of Things}
\acrodef{CAD}{Computer-Aided Design}
\acrodef{EDA}{Electronic Design Automation}
\acrodef{HPC}{High-Performance Computing}
\acrodef{DL}{deep learning}
\acrodef{ML}{machine learning}
\acrodef{NLP}{natural language processing}
\acrodef{IC}{Integrated Circuit}
\acrodef{CWE}[CWE]{Common Weakness Enumeration}
\acrodef{CVE}[CVE]{Common Vulnerabilities and Exposures}
\acrodef{LLM}[LLM]{large language model}
\acrodef{NMT}[NMT]{neural machine translation}
\acrodef{IP}[IP]{hardware intellectual property block}
\acrodef{HDL}[HDL]{hardware description language}
\acrodef{RTL}[RTL]{register-transfer level}
\acrodef{SDL}[SDL]{security development lifecycle}
\acrodef{FSM}[FSM]{finite state machine}
\acrodef{AST}[AST]{abstract syntax tree}
\acrodef{SoC}[SoC]{system-on-chip}

%% file: section/preamble/macros.tex
\newcommand{\ignore}[1]{{}}

\newcommand{\squishlist}{
	\begin{list}{$\bullet$}
		{ \setlength{\itemsep}{0pt}
			\setlength{\parsep}{1pt}
			\setlength{\topsep}{1pt}
			\setlength{\partopsep}{0pt}
			\setlength{\leftmargin}{0.9em}
			\setlength{\labelwidth}{1.5em}
			\setlength{\labelsep}{0.4em} } }
	\newcommand{\squishend}{
	\end{list}  }

\definecolor{graphFirst}{RGB}{2,136,209} 
\definecolor{graphSecond}{RGB}{211,47,47} 
\definecolor{graphThird}{RGB}{245,124,0} 
\definecolor{graphFourth}{RGB}{56,142,60} 
\definecolor{graphFifth}{RGB}{81,45,168} 
\definecolor{graphSixth}{RGB}{69,90,100} 
\definecolor{graphSeventh}{RGB}{251,192,45} 
\definecolor{backgroundSecond}{RGB}{239,154,154} 
\definecolor{backgroundThird}{RGB}{255,204,128} 
\definecolor{backgroundFourth}{RGB}{165,214,167} 
\definecolor{backgroundFifth}{RGB}{179,157,219} 
\definecolor{backgroundSixth}{RGB}{176,190,197} 
\definecolor{backgroundSeventh}{RGB}{255,245,157} 

\newcommand{\Com}[1]{}

%% file: section/01Introduction.tex
\section{Introduction\label{sec:intro}}

The cost of correcting errors in hardware designs increases considerably in later stages of the product development cycle. 
Despite stringent requirements on functional correctness~\cite{tasiran_coverage_2001}, hardware bugs, including security bugs, can persist~\cite{dessouky_hardfails_2019}. 
A primary reason for this stems from the difficulty in evaluating security.
Defining security properties and evaluating them is a complex process, requiring rare domain expertise.
While several tools exist to support designers, such as linting tools~\cite{vclint_synopsys_2022}, they often focus on checking functional correctness and only have limited features out-of-the-box for verifying security-related properties. 
Such tools can be used for some security-related checking~\cite{dessouky_hardfails_2019, bidmeshki_hunting_2021} but considerable security-relevant expertise is required. 
These tools are designed to function on near-complete RTL code, performing comprehensive and time-consuming testing to ensure all aspects of a given system are compliant. 
However, improving security requires checks throughout the  design cycle, including in the early stages of RTL design where a complete security specification is not yet available.

Recently, industry-led efforts have added hardware-related issues to the \ac{CWE} list that is hosted by the MITRE Corporation~\cite{the_mitre_corporation_mitre_cwe-1194_2021}. 
The presence of a given weakness indicates that there are ``flaws, faults, bugs, or other errors in software or hardware implementation, code, design, or architecture that if left unaddressed could result in systems, networks, or hardware being vulnerable to attack''~\cite{the_mitre_corporation_mitre_cwe-1194_2021}. \acp{CWE} serve as a ``common language'' for navigating weaknesses. 
Identifying different weaknesses requires different levels of \textit{context} about the design, designers' intent, and security policies. 
To the best of our knowledge, \textbf{early-stage} analyses of hardware-based security weaknesses are largely a manual effort~\cite{dessouky_hardfails_2019} that includes human inspection of \ac{HDL} code~\cite{bidmeshki_hunting_2021}. 
Ideally, this process of code inspection should be supported through automated scanning, with feedback to designers as they go. 
This analysis subsequently complements other later-stage efforts related to simulation/testing (e.g.,~\cite{meng_rtl-contest_2021}), and formal verification (e.g.,~\cite{ray_invited_2019,nienhuis_rigorous_2020}) as part of an overall \ac{SDL} (e.g.,~\cite{dorsey_intel_2020}). 

In this work, we explore the practical implications and feasibility of automating the detection of hardware \acp{CWE}, focusing on a selection of weaknesses with minimal context. 
We investigate a set of \acp{CWE} as a case study, and we present our experience in implementing static analysis scanners to detect them, providing insights into the process of defining, implementing, and refining \ac{CWE} scanners. 
Following a discussion of the motivation for our work (\autoref{sec:background-and-related-work}), our main contributions are:
\begin{itemize}
    \item We investigate the hardware \ac{CWE} database for weaknesses that are potentially amenable to identification from source code analysis (\autoref{sec:cwe-classification}).
    \item We propose and prototype scanners to provide security-related feedback (\autoref{sec:algorithms}).
    \item We evaluate our proposed scanning algorithms on a series of open-source SoC and processor designs (\autoref{sec:results}) and discuss their limitations and other practical issues (\autoref{sec:dis}).
\end{itemize}

%% file: section/01x_Background.tex
\section{Background\label{sec:background-and-related-work}}

\subsection{Related Work}
There are several approaches to improving the security of hardware designs, including the use of a \acf{SDL}~\cite{dorsey_intel_2020} that runs in parallel with traditional functionality driven development. 
%
The \textbf{Planning} stage is the first stage for \ac{SDL} where security requirements are specified. 
This is followed by the \textbf{Architecture} and \textbf{Design} stages, where relevant threat models are considered. 
Checking the design against these security threat models is carried out in the \textbf{Implementation} and \textbf{Verification} stages. 
While \textbf{Implementation} consists of manual checks and static analysis of code, the bulk of validation is carried out in \textbf{Verification} where security properties expressed through assertions in \acp{HDL} are verified. Physical testing is also done after fabrication. 
The last stage is \textbf{Release and Response}.

Several security analysis techniques can be used throughout this process. 
Recent work has proposed formal verification \cite{bloem_formal_2018, he_soc_2019, nienhuis_rigorous_2020, ray_invited_2019, sepulveda_towards_2018}, Information Flow Tracking \cite{ardeshiricham_register_2017, brant_challenges_2021, hu_detecting_2016, zhang_hardware_2015}, fuzz testing \cite{tyagi_thehuzz_2022, trippel_fuzzing_2021, muduli_hyperfuzzing_2020, accellera_systems_initiative_draft_2021}, as well as run-time detection \cite{hicks_specs_2015, mushtaq_whisper_2020, sarangi_phoenix_2006}. 
These techniques rely on simulation or they operate in the field, using complete or near-complete designs. 
To complement these, we focus on an earlier stage in the design process, specifically in the \textbf{implementation} phase. 
Our premise is that \textbf{static analysis of \ac{RTL} code} can identify some security weaknesses, preventing them from propagating to the next stage. 

Static analysis is an established approach used in software to identify errors and bugs; e.g., \texttt{Lint} was proposed in 1978 to analyze C code~\cite{johnson_lint_1978}, and has since been adopted in many programming languages coining the term "Linter". 
Lint algorithms try to strike a balance between accuracy and practicality, as generated warnings are acceptable only in proportion to the real bugs uncovered. 
If too many false positives are raised, they obscure true problems. 




Several commercial and community tools offer linting capability for HDL (e.g,~\cite{vclint_synopsys_2022,noauthor_verilator_2021}), but they focus predominantly on functional and structure checks. 
They are comprehensive tools with several tags and parameters that can be configured to check for custom and predefined rules. 
Linting has been used to handle critical hardware issues like latches and clock gate timing validation \cite{soliman_automatic_2018}. 
Formal linting tools have been analyzed to find bugs in \ac{RTL} before synthesis \cite{yadav_study_2020}. 
The idea of continuous linting is proposed in \cite{hansson_continuous_2014} to bring design errors to the attention of designers at a faster rate. 
While these works highlight the capability of linting for reducing the load on the verification stage, they do not look at hardware that could be functionally correct yet insecure. 
It is up to the security expertise of the designer to identify a lint message relevant to security.

\subsection{Motivation}
Consider the following scenario: an \ac{RTL} designer works on several hardware modules, iterating the design and periodically checking it via simulation, synthesis, or manual review. 
As they work, they require frequent, fast feedback -- for example, a simulation tool will report syntax errors and potentially help the designer avoid design bugs. 
For a more thorough check, the designer could use some kind of static analysis tool, such as a \textbf{linter} to check the source code against set guidelines~\cite{bening_principles_2001} and raise warnings or error messages. 

To explore the potential limitations of linting tools, we investigated an industry tool (Tool\#1), and Verilator~\cite{noauthor_verilator_2021}. We set Verilator in \texttt{lint-only} mode with the \texttt{-Wall} flag (enables all checks and warnings) set.
We set Tool\#1 on the \textbf{lint\_rtl} goal under the \textbf{block/initial\_rtl} methodology, appropriate for early-stage \ac{RTL} design. 
After analyzing the hardware \acp{CWE} for those that are amenable to static analysis (discussed in~\autoref{sec:cwe-classification}), we implemented a small set of Verilog modules with weaknesses based on the examples on the MITRE website. 
The results are in \autoref{tbl:lint-comparison-table}.
The default set of rules for Tool\#1 could not find security-specific issues.
However, as Tool\#1 offers optional rules, it might be possible to find security-relevant ones. A designer must identify security-relevant tags in the 936 available rules, a challenge!
We identified seven rules covering \ac{FSM}-related problems, race conditions, and uninitialized registers that could lead to security bugs. 

We ran Tool\#1, adding the identified tags to the default \ac{RTL} lint rules, on the HACK@DAC'21 SoC~\cite{hackevent_hackdac21_2022,dessouky_hardfails_2019} and it raised 597 warnings in 84 seconds. 
Only 19 of the warnings were related to the optional tags with the rest raised by the default rules, representing considerable ``noise,'' i.e., issues not specific to security, although a designer can reduce noise by creating a custom goal with only the security-specific linting rules. 
With this setting, the linter was only able to detect potential issues related to state encoding in \acp{FSM}. 
We expected to detect \ac{FSM} related issues with Tool\#1 but they were not identified in our benchmarks. This probably happens because our implementation of \acp{FSM} does not follow standard \ac{FSM} related patterns.  
Another plausible explanation is that basic lint checks with industry tools require the elaboration of the design and do not obtain all relevant information needed at the earlier analysis stage. The details of these stages are discussed in the next section.
All in all, \textbf{linters are not designed specifically with security in mind} and require time and expertise to set them up for security-related tasks and to filter the results. 
We are thus motivated to investigate the feasibility of identifying security bugs with pre-elaboration static analysis to provide early-stage security feedback to designers.

\begin{table}[h]
\footnotesize
\caption{Out-of-the-box linting for security}
\vspace{-1em}
\label{tbl:lint-comparison-table}
\resizebox{\linewidth}{!}{
\begin{tabular}{@{}lllccc@{}}
\toprule
CWE         & \begin{tabular}[c]{@{}l@{}} Module \end{tabular} & \begin{tabular}[c]{@{}l@{}}\ Weakness \end{tabular} & \begin{tabular}[c]{@{}c@{}}Tool\#1 \end{tabular} & \begin{tabular}[c]{@{}c@{}}Verilator \end{tabular} & \begin{tabular}[c]{@{}c@{}}\sol{} \end{tabular}                 \\ \midrule
1234 &  Locked register     & debug overrides lock                &  \redtxt{\xmark} & \redtxt{\xmark} & \greentxt{\cmark} \\
1271 &  JTAG lock           & jtag lock signal not reset          &  \redtxt{\xmark} & \redtxt{\xmark} & \greentxt{\cmark} \\
1245 &  state machine       & incomplete case statement           &  \redtxt{\xmark} & \greentxt{\cmark} & \greentxt{\cmark} \\
1245 &  state machine       & unreachable state                   &  \redtxt{\xmark} & \redtxt{\xmark} & \greentxt{\cmark} \\
1245 &  state machine       & FSM deadlock                        &  \redtxt{\xmark} & \redtxt{\xmark} & \greentxt{\cmark} \\
1280 &  Access control      & access set after transfer           &  \redtxt{\xmark} & \redtxt{\xmark} & \greentxt{\cmark} \\
1262 &  Peripheral sensor   & write signal not checked            &  \redtxt{\xmark} & \redtxt{\xmark} & \greentxt{\cmark} \\
\bottomrule
\end{tabular}}
\vspace{-2.4em}
\end{table}

%% file: section/02StaticAnalysisCWEs.tex

\section{Static Analysis of Hardware CWEs}
\label{sec:cwe-classification}

This work aims to improve security-specific analysis and feedback in early-stage \ac{RTL} design by finding hardware \acp{CWE}.
However, the \ac{CWE} list comprises myriad weaknesses at different levels of specificity and abstraction.
We first classify the \acp{CWE} based on our assessment of how amenable they are to static analysis, and then propose the framework for our \unbf{CWE} \unbf{A}nalysis \unbf{T}echniques, \sol{}. 


\subsection{Classifying CWEs for Scanning}

There are 96 HW \acp{CWE} on MITRE's website~\cite{the_mitre_corporation_mitre_cwe-1194_2021} at the time of writing.
In this section, we classify them by their amenability to detection, with our classification of the \acp{CWE} presented in~\autoref{tbl:classification-CWEs}.

The first delineation is whether the \ac{CWE} is code-based or not. We exclude \acp{CWE} that do not originate in \ac{RTL} designs (6).
We consider whether the \acp{CWE} can be detected statically or dynamically (i.e. in a simulation or at run-time). Our approach focuses on \acp{CWE} that can be statically detected. \acp{CWE} such as CWE-1264, where de-synchronization between data and permission checking logic is evident only at run-time, require a different approach for detection (30 require functional simulation). We examine the \acp{CWE} that can be detected statically, classifying them based on the data required.   

Some bugs may be detectable using only \textbf{static analysis} of the \ac{RTL} code, where the source is transformed into parse-trees, performing type-inferencing and detecting syntax errors. Here, the created parse-trees can be further analyzed using different algorithms to extract useful information for \ac{CWE} identification. These can be classified into those that are directly detectable from the code (11), or those that may require additional context (21).

Other \acp{CWE} can be further partitioned by the amount of ``extra information'' or \textit{context} required to scan for a weakness (9 \acp{CWE} total).
\textbf{Static elaboration} of the design entails binding instances to modules, resolving library references, processing \textit{defparam} statements, unrolling \textit{for} loops, flattening instance arrays, and replacing parameters with constants. 
The result after static elaboration is an elaborated parse tree.
\textbf{RTL elaboration} then synthesizes the design using pre-defined primitives (e.g., AND, OR) and operators (e.g., add, multiply). 
The result of \ac{RTL} elaboration is a synthesized \ac{RTL} netlist.
The remaining 16 CWEs do not have general symptoms among designs, so they cannot be formalized and identified, requiring manual analysis by IP security experts to detect.

As an example of the different types, consider the following. 
An incomplete \ac{FSM} can be discovered by static analysis without any information from users (CWE-1245). 
In contrast, to identify the improper translation of security attributes by the fabric bridge (CWE-1311), additional information (context) regarding the fabric protocol is required. 
Even though several \acp{CWE} can be detected by static analysis, some issues do not manifest until the parse-tree is elaborated. 
For instance, incorrect default module parameters (CWE-1221) are difficult to find by static analysis. 
However, we can identify this issue through static elaboration, where all the parameters are calculated and propagated to each module. 
Moreover, cross-modular issues might demand \ac{RTL} elaboration, e.g., CWE-1291. 
Still, some \acp{CWE} cannot be discovered at the \ac{RTL} design stage, e.g., CWE-1278, which requires reverse-engineering protection. 

\begin{table}[t]
\caption{Classification of Hardware CWEs}
\label{tbl:classification-CWEs}
\vspace{-1em}
\resizebox{\linewidth}{!}{%
\begin{tabular}{@{}llc@{}}
\toprule
Detection method & CWEs & \# CWEs \\ \midrule

Non-RTL  & 1053, 1059, 1261, 1273, 1248, 1266    & 6      \\\hline

Functional simulation & \begin{tabular}[c]{@{}l@{}}1342, 1296, 1246, 1303, 1193, 1279, 1281, 1323, 440, 1264, 1297\\ 1191, 1258, 1272, 1313, 1324, 1223, 1224, 1233, 1263, 1251, 1315\\ 1304, 1332, 441, 1268, 1294, 1190, 203\end{tabular} & 29      \\\hline

Static analysis (no context)  & \begin{tabular}[c]{@{}l@{}}1234, 1245, 1262, 1271, 1280, 1244, 1231, 226, 1282, 1232, 1314\end{tabular} & 11  \\\hline

Static analysis (with context)& \begin{tabular}[c]{@{}l@{}}1318, 1291, 1311, 1312, 1302, 1326, 1310, 1209, 1269, 1257, 1316, \\1317, 1320, 1189, 1242, 1334, 1274, 1283, 1243, 1295, 1240, 1328\end{tabular} & 22  \\\hline

Static/RTL elaboration & 1254, 1298, 1276, 1331, 1220, 1221, 276, 1260, 1299  & 9  \\\hline

Manual analysis      & \begin{tabular}[c]{@{}l@{}}1252, 1253, 1192, 1277, 1267, 325, 1241, 1256, 1300, 1301, 1278\\1319, 1255, 1338, 1351, 1247\end{tabular}& 16       \\ \bottomrule

\end{tabular}
}
\vspace{-1.5em}
\end{table}

In this work, we focus on static analysis of \ac{RTL} source per-module with little to no context. 
Thus, \acp{CWE} that are specific to limited module-types, cross-modular \acp{CWE}, or those requiring simulation are out of scope and are the focus of our future work. 
We target the following five \acp{CWE} as representatives for illustration. 

\noindent\textbf{CWE-1234:} Hardware Internal or Debug Modes Allow Override of Locks. A debug or debug-related signal overrides the value of an internal security-relevant signal, e.g., a lock or access control signal. This allows adversaries to reconfigure sensitive policy values or disable security protections if they can get into debug mode.

\noindent\textbf{CWE-1271:} Uninitialized Value on Reset for Registers with Security Settings. Registers with sensitive information like keys, lock bits or access control bits do not default to secure postures on reset. 

\noindent\textbf{CWE-1245:} Improper Finite State Machines (\acp{FSM}) in Hardware Logic -- where \ac{FSM} logic contains ``gaps'' which could allow adversaries to put the system in an unknown or vulnerable state. 

\noindent\textbf{CWE-1280:} Access Control Check Implemented After Asset is Accessed -- where an asset is accessible before the check on its access control signal. The asset should be accessible only after the check is successful, otherwise the security might be compromised.

\noindent\textbf{CWE-1262:} Improper Access Control for Register Interface -- where a register value is not write-protected or is not correctly protected, i.e., untrusted users can override the signal through the register interface. The protection mechanisms for security-relevant signals must be correctly implemented to avoid data leakage.

\subsection{CWEAT Framework}
\begin{figure}[b] 
    \centering
    \includegraphics[width=0.8\linewidth]{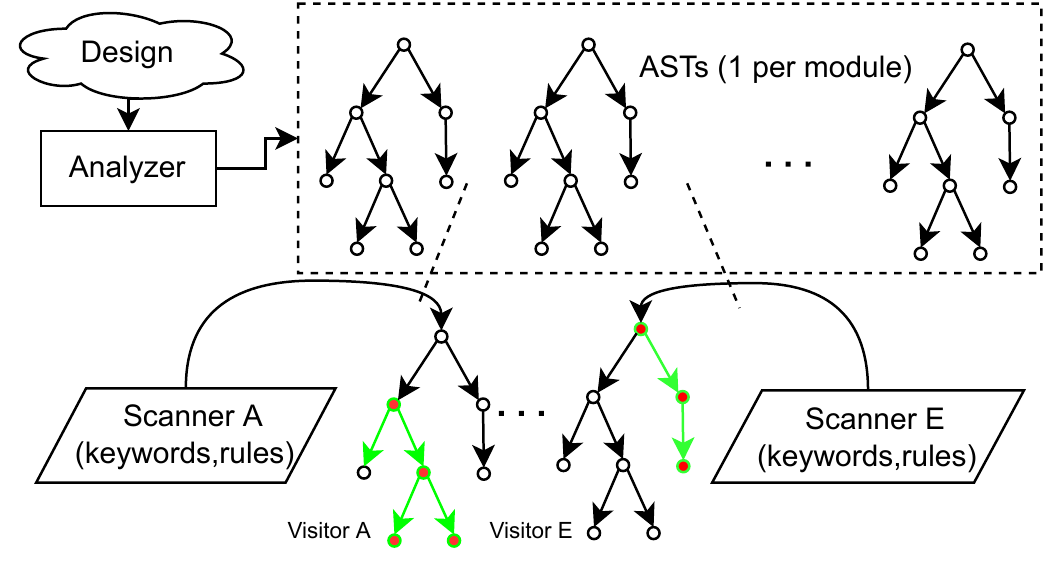}
    \vspace{-3mm}
    \caption{\sol{} Framework. \acp{AST} are extracted from the design and fed through CWE scanners}
    \label{fig:cweeet-framework}
\end{figure}

To assist in the detection of our targeted \acp{CWE}, we propose to create \acp{AST} from Verilog source and traverse them with hand-crafted, heuristic detection algorithms (\autoref{sec:algorithms}). 
\autoref{fig:cweeet-framework} illustrates the \sol{} framework, where the input is the design repository containing the source code. 
The design can comprise a single \ac{IP} through to an entire \ac{SoC}. 
Relevant source code files (in Verilog and/or SystemVerilog) are extracted from this repository and sorted in the order they should be analyzed. 
If a file has errors or does not have necessary dependencies, the tool ignores it. 
The final sorted source code list is sent to the Analyzer. 
The Analyzer runs on the entire set of source code files and produces $N$ \acp{AST} -- one per module. 
We use Verific~\cite{noauthor_verific_2022} as the front-end parser to build the \acp{AST}. 
Each scanner traverses every \ac{AST} according to the detection algorithm for a \ac{CWE} using the associated rules and keywords. 
Future work can add new scanners to the framework. 

We use keywords while searching for weaknesses for two reasons: (i) to locate `security relevant' features in RTL (discussed in detail in \autoref{sec:algorithms}), and (ii) to prune the search space for these weaknesses. 
The keyword-matching rules comprise matching and exclusion rules, e.g., ignoring \texttt{clock} when looking for \texttt{lock}. 
These rules could come from established design styles or might be provided by the designers. 
For our experiments, we devised keyword-matching rules by  observing design styles. 

\Com{
\todoblock{Explain the security design life-cycle? Other background material? Maybe a brief overview of dynamic vs. static analysis, and why we want to go in this direction (move II-D)? Outline research questions -- lead to the CWEAT tool not as a "tool" being the contribution, but as the vehicle for our research investigation into the feasibility of the earlier-stage feedback. $\rightarrow$ Background and Related Prior Work}

\todoblockblue{Can we come up with a quantitative way to show that this program is hard? State space explosion explanations? Search space?}

\todoblock{What is the argument for doing the scanning here? Why hasn't this been tried before? The Hack@DAC results/retrospective papers are probably going to be very important here...? Why not use software linting?}
}

\Com{ 
\textbf{Motivation}
Linting tools were first proposed in 1978 as a static code analysis tool to identify errors and bugs in C \cite{Johnson1978LintAC}. They have since been adopted in any programming language and eventually made their way in \acp{HDL} too. It was noted from the beginning how lint algorithms are a compromise between accuracy and practicality, and how the generated warnings are acceptable only in proportion to the real bugs uncovered. If too many false positives are raised, they will only obscure the true problems. In this optic we tried to see which security bugs could be detected using VC SpyGlass Lint, an industry grade linter which offers a huge set of optional rules. We imagined the scenario in which a designer wants to use this tool to uncover security bugs early on during the design process. This designer is experienced but with limited knowledge about security specific topics. The first problem that is encountered is that the default set of rules does not cover any security specific issue. The designer will have to identify security relevant tags in the 936 available ones. We were able to identify 5 of them covering \ac{FSM} related problems and racing conditions that might lead to security bugs. The second issue that we encounter is that even basic language checks require the elaboration of the design, implying that this process cannot be run while writing the design like software linters. We ran VC SpyGlass Lint adding the identified tags to the default lint rules on the HACK@DAC21 SoC and it raised 597 warnings in 84 seconds. Only 19 of the warnings were related to the custom tags with the rest raised by the default rules. The designer could reduce the noise down by creating a custom goal with only the specific linting rules. The linter was only able to detect potential issues relelated to state encoding. All in all, linters have not been designed with security in mind and require time and expertise to set them up for this tasks and to filter the results, while still being able to identify just a very small set of potential security problems. With CWEAT we investigate the feasibility of identifying security bugs with pre-elaboration static analysis to provide early-stage feedback to designers.
}
\Com{
\todoblockblue{Is it possible to quantify the "savings" from using this approach? Compare to random search? Human search?}
}

\Com{
From the analysis of static-analyzable \acp{CWE} and playing with linting tools we learned the following lessons:
\begin{itemize}
    \item Some linting rules are too general, causing too many warnings unrelated to security;
    \item Some CWE-specific rules might be too complex for an exhaustive analysis.
\end{itemize}
For this reason we adopted some keyword-matching rules that allow us to reduce noise in the raised warnings and reduce the search space to get fast performances. We identified a set of keywords that help us identify the security related signals to be considered in the analysis for \acp{CWE} detection. The keywords-matching rules are actually composed of matching and unmatching rules to avoid for example to match the word \texttt{clock} when looking for \texttt{lock}.
The considered keywords and their frequency in the considered SoC and cores designs is reported in Figure \ref{fig:keywords}.
}


\Com{ 
\subsection{\rt{Keyword search justification}}
\label{sec:CWEAT-keywords}

\todoblock{Explain the motivation or need for tools that can narrow the search space, especially if manual inspection is generally the main way to get things done~\cite{dessouky_hardfails_2019}. Also argue that we intuit that designers are often sensible and will use reasonable names to describe their signals and modules (hook into the discussion section the possibility of generalizing further? Or explain how we started in the more general case, but inevitably we will need to reduce the findings down. }

\begin{table}[!ht]
    \centering
    \begin{tabular}{llll}
    \toprule
        Design & \# Sig. & \# Sig. O.I & \# Mod\\ \midrule
        H@D21 & 7619 & 1533 & 312  \\ 
        H@D18 & 9250 & 2671 & 407  \\ 
        e203 & 11881 & 3190 & 141  \\ 
        orpsoc & 8716 & 2737 & 185 \\ 
        ibex & 2071 & 479 & 34     \\ 
        cv32 & 2314 & 534 & 107    \\ 
        SCR1 & 5185 & 1620 & 75    \\ 
        serv & 304 & 109 & 15      \\ 
        ultraemb. & 651 & 143 & 17 \\ 
        mor1kx & 2742 & 363 & 46   \\ \midrule
        TOT & 50733 & 13379 & 1339 \\ \bottomrule
    \end{tabular}
\end{table}
}

%% file: section/03Algorithms.tex
\section{CWE Detection Algorithms\label{sec:algorithms}}

\subsection{Preliminaries}
\begin{figure}[htb]
    \centering
    \begin{subfigure}[b]{0.95\linewidth}
        \begin{lstlisting}[language=verilog]
module sum(input wire [7:0] a, output [7:0] out);
    always @(a) begin
        out = a + 1;
    end
endmodule 
        \end{lstlisting}
        \vspace{-1mm}
    \end{subfigure}
    
    \begin{subfigure}[b]{0.9\linewidth}
        \centering
        \includegraphics[width=\linewidth]{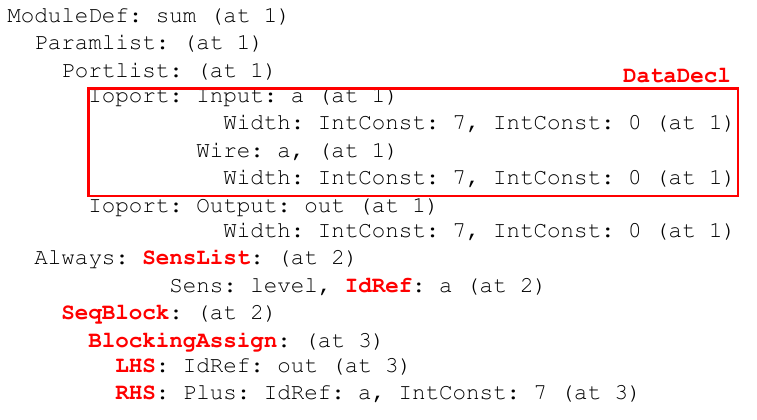}
    \end{subfigure}
    \vspace{-3mm}
    \caption{Example Abstract Syntax Tree of Verilog module}
    \label{fig:ast_ex}
\end{figure}

Our heuristic detection algorithms work on the \acp{AST} of an \ac{HDL} design\footnote{Examples use a Verilog-like syntax but the concepts are applicable to all \ac{HDL}s.}. The \ac{AST} can be extracted by a parser of choice for the preferred \ac{HDL}. Each node of the \ac{AST} represents a construct defined by the formal grammar of the \ac{HDL}, e.g., a declaration, an expression or a module instance. The nodes are connected together to preserve the syntax expressed in the parsed design. \autoref{fig:ast_ex} shows an example of the \ac{AST} of a Verilog module.

A common design pattern to traverse \acp{AST} is the Visitor Pattern~\cite{gamma_design_1995}. 
This pattern involves defining a procedure that will be executed for each construct of interest. 
When a node is traversed, if a procedure for the  construct exists, it is executed; otherwise, the children nodes are traversed. 
This allows for a built-in recursion that ends at leaf nodes or at specified constructs for which the children nodes are not traversed. 
A visitor accepts a node $N$ when it starts the traversal of the sub-tree with $N$ as root.

\subsection{CWE Scanner Algorithms}

\subsubsection{Scanner-1234: }

\begin{algorithm}
\caption{: CWE-1234 Detector}\label{alg:1234}
\begin{algorithmic}

\Procedure{traverse}{$ConditionalStatement$ $node$}

    
    \If{$node.if\_expr.matches(lock\_kw,dbg\_kw,ops)$}
        \State $results.append( Result(if\_expr, loc) )$
    \EndIf
    
    \State TRAVERSE$(node.ElseStmt)$, TRAVERSE$(node.ThenStmt)$
    
\EndProcedure

\end{algorithmic}
\end{algorithm}

\autoref{alg:1234} illustrates the scanner for \ac{CWE}-1234. 
As the tree is traversed, the expression for the \texttt{if} statement in \texttt{ConditionalStatement} node(s) is fetched as \texttt{if\_expr}. 
\texttt{if\_expr} is analyzed by checking whether it matches three conditions: (i) presence of a lock keyword from a list \texttt{lock\_kw}, (ii) presence of a debug keyword from a list \texttt{dbg\_kw}, and (iii) presence of binary logic operators (e.g., \texttt{or}) that indicate `overriding'. 
If these conditions are met, the \texttt{if\_expr} is categorized as potentially vulnerable. 
Relevant information is collected and appended to \texttt{results}.
Next, the \texttt{else} and \texttt{then} statements of the \texttt{ConditionalStatement} are traversed sequentially. 
This recursion takes care of nested if-else statements. 



\subsubsection{Scanner-1271: }

\begin{figure}[tbp]
        \centering
        \includegraphics[width=0.85\linewidth]{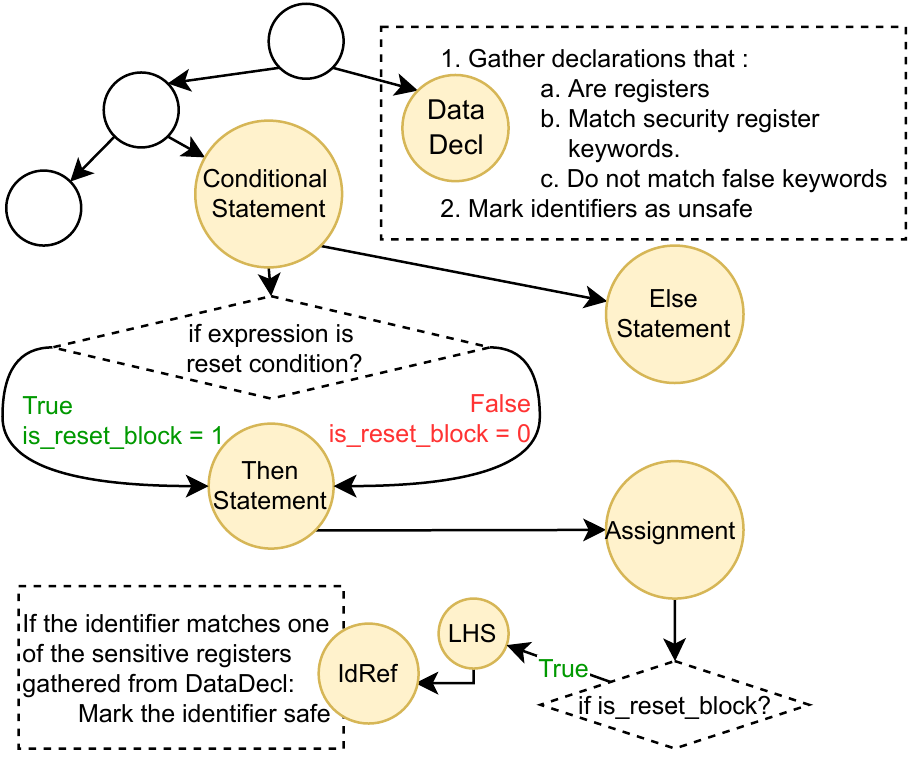}
        \vspace{-3mm}
        \caption{Traversal for \ac{CWE}-1271}
        \label{fig:1271-alg}
\end{figure}


\autoref{fig:1271-alg} illustrates the scanner for \ac{CWE}-1271. 
While traversing a data declaration node, candidates for security relevant registers are gathered and marked as unsafe. 
These candidates are then marked as safe based on the rest of the traversal. 
In RTL, this happens if the security-relevant register is assigned a value inside a reset block. 
In the \ac{AST}, this is identified through traversing various nodes based on flags set in the scanner. 
The flag \texttt{is\_reset\_block} is set when traversing a node which represents a reset block. 
The flag \texttt{is\_reset\_lhs} is set when traversing a node that is the left hand side of an assignment inside a reset block.

\begin{figure}[t]
    \begin{subfigure}{.25\textwidth}
        \centering
        \includegraphics[width=.6\linewidth]{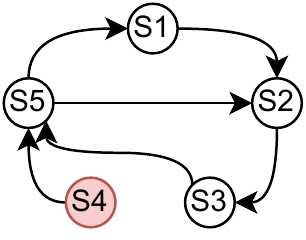}
        \vspace{-1mm}
        \caption{Unreachable state (S4)}
        \label{fig:fsm-weaknesses-1}
    \end{subfigure}%
    \begin{subfigure}{.25\textwidth}
        \centering
        \includegraphics[width=.6\linewidth]{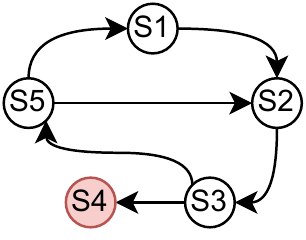}
        \vspace{-1mm}
        \caption{\ac{FSM} deadlock}
        \label{fig:fsm-weaknesses-2}
    \end{subfigure}
    \vspace{-3mm}
    \caption{\ac{FSM} weaknesses}
    \label{fig:fsm-weaknesses}
    \vspace{-1.42em}
\end{figure}

\subsubsection{Scanner-1245: }

has a 3-step detection strategy:\\
    \noindent(I) \textit{Find \ac{FSM} variables:} The scope of the module, containing the locally declared identifiers in a symbol table, is obtained. 
    The identifiers are checked to be part of explicit FSMs and of the type \texttt{Variable}. Explicit FSMs are easily recognize as one-hot encodable and are modeled with a single clocking event. Implicit FSMs are found similarly after being converted to explicit FSMs.
    All variables linked to this \ac{FSM} variable are tracked and added to a list of variables belonging to a specific FSM.\\
    \noindent(II) \textit{Extract \ac{FSM} transitions:} This traversal progresses according to the variables found in the previous step and gathers information from assignments of these variables. This information includes the previous state, next state, condition for transition and the location of assignment. The goal here is to collect the transitions of an \ac{FSM} and add them to the \ac{FSM} object.\\
    \noindent(III) \textit{Analyze FSM:} After relevant information is gathered, the \ac{FSM} is checked for several weaknesses. These include unreachable states, \ac{FSM} deadlocks, and incomplete case statements. An unreachable state is shown as \texttt{S4} in \autoref{fig:fsm-weaknesses-1}; it is the initial state of a transition, but there is no way to get there, i.e., \texttt{S4} is not present in the final state of any transition or on reset. An \ac{FSM} deadlock is shown as \texttt{S4} in \autoref{fig:fsm-weaknesses-2}; it is the final state of a transition, but there is no way to get out i.e., it is not the initial state of any transition. A case statement representing an \ac{FSM} is analyzed for completeness in addition to these \ac{FSM} vulnerabilities. Completeness is verified if there is either a default statement present or if the number of case items equals the possibilities from the size of case condition.


\subsubsection{Scanner-1280: }

\Com{
\begin{algorithm}
\caption{CWE-1280 Detector}\label{alg:1280}
\begin{algorithmic}


\Procedure{traverse}{$AlwaysConstruct \: node$}
    \State $v.sens\_list \gets node.sens\_list$
    \State TRAVERSE$(node.Stmt)$
    \State $v.sens\_list \gets [\hskip.25em]$
\EndProcedure
\State  \vspace{-5pt}
\Procedure{traverse}{$SeqBlock \: node$}
    \State SeqBlock $lastSeqB$
    \State $lastSeqB \gets v.last\_block$
    \State $v.last\_block \gets node$
    \State TRAVERSE$(node.GetStmts)$
    \State $v.last\_block \gets lastSeqB$
    
\EndProcedure
\State \vspace{-7pt}
\Procedure{traverse}{$BlockingAssign \: node$}
    \State $v.is\_b\_assign \gets True$
    \State $v.is\_lhs \gets True$
    \State TRAVERSE$(node.GetLhs)$
    \State $v.is\_lhs \gets False$
    \State TRAVERSE$(node.GetRhs)$
    \State $v.is\_b\_assign \gets False$

\EndProcedure
\State \vspace{-7pt}
\Procedure{traverse}{$IdRef \: node$}

    \If{$v.is\_b\_assign$ \textbf{and} $node.id$ \textbf{not in} $v.sens\_list$}
        \State $t \gets Id(node.id, v.last\_block, node.lineno)$
        \If{$v.is\_lhs$}
            \State $v.writes.add(t)$
        \Else    
            \State $v.reads.add(t)$
        \EndIf
    \EndIf
    
\EndProcedure
\State  \vspace{-2pt}
\Procedure{Detect\_1280}{$Module \: mod$}
    \State Visitor1280 $v$
    \State $mod.Accept(v)$
    \For{$id\_w, b\_w, lno\_w$ \textbf{in} $v.writes$}
        \For{$id\_r, b\_r, lno\_r$ \textbf{in} $v.reads$}
            \If{$id\_r=d\_w$ \textbf{and} \\ $b\_r=b\_w$ \textbf{and} \\ $lno\_r<lno\_w$}
                \State $LogCWE(1280, lno\_r, ln\o_w)$
            \EndIf
        \EndFor
    \EndFor
\EndProcedure

\end{algorithmic}
\end{algorithm}
}

\begin{figure}[h] 
    \centering
    \includegraphics[width=1\linewidth,height=3.5cm]{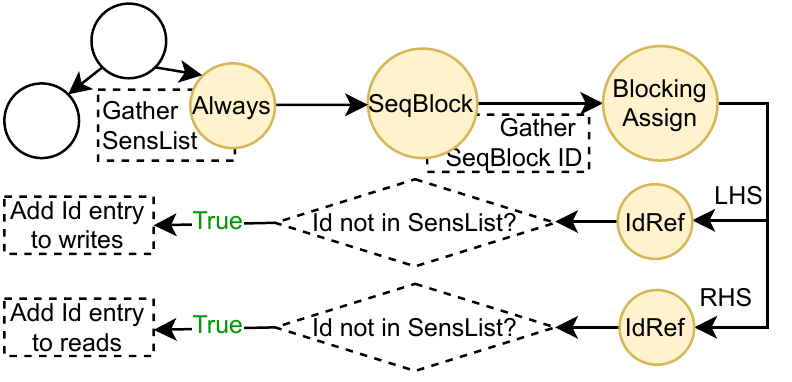}
    \vspace{-3mm}
    \caption{Traversal for \ac{CWE}-1280}
    \label{fig:1280-alg}
    \vspace{-1.42em}
\end{figure}

\autoref{fig:1280-alg} illustrates the scanner for \ac{CWE}-1280. 
A traversal of the \ac{AST} builds two lists containing information on read and write operations (\texttt{reads} and \texttt{writes}) lists respectively). 
The entries of the two lists are triples containing the \texttt{Id}, a reference to its parent \texttt{SeqBlock}, and its line number. 
After visiting the \ac{AST} of the module  and building the \texttt{reads} and \texttt{writes} lists, the algorithm iterates through these lists and compares each read entry with the write entries. If a read and a write have the same ids, are in the same \texttt{SeqBlock}, and if the read comes before the write, the updated value is not read, posing a potential weakness. 

\Com{, the procedures are presented in Algorithm~\ref{alg:1280}. A Visitor class Visitor1280 is used to build two lists containing information read and write operations ($reads$ and $writes$ lists respectively) which are then analyzed to detect the CWE. Two flags, $is\_lhs$ and $is\_b\_assign$ are used to signal that the traversed node is inside a left-hand side (lhs) of an assignment and inside a blocking assign respectively. $last\_block$ keeps track of the last visited begin-end block. $sens\_list$ stores the ids of the sensitivity list of the parent always block, is an empty list if the current node does not have a parent always block. 

When traversing an always construct the visitor first copies the sensitivity list into $sens\_list$ then proceeds with the traversing of the statement. Before returning, it clears the $sens\_list$ by assigning it to an empty list.

When traversing a sequential block, the visitor stores the last visited sequential block, $v.last\_block$, in a temporary variable, $lastSeqB$, then sets $v.last\_block$ to the current node before proceeding the traversing on the block statements. before returning it restores $v.last\_block$ to the original value assigning it $lastSeqB$.

When traversing a blocking assign, the visitor sets the flags $v.is\_b\_assign$ and $v.is\_lhs$ to $True$, then it visits the lhs of the assignment. After the visit to the lhs, it sets the flag $v.is\_lhs$ to $False$ and proceeds traversing the right-hand side. Before returning it sets the flag  $v.is\_b\_assign$ to $False$.

When traversing an id reference, the visitor checks if this id is in a blocking assign and is not in the sensitivity list of the parent always block. If this condition is met, then a tuple with the information about the id is created and then added to the $writes$ list if the id is in a lhs, to the $reads$ list otherwise.
}

\subsubsection{Scanner-1262:}
\label{sec:1262}
has a 2-step detection strategy:\\
    \noindent(I) \textit{Find the control signals for the registers associated with write transactions:} 
    We only focus on registers that are not properly `write protected', so only signals with names containing \texttt{wdata} are considered.
    As the \ac{AST} is traversed, we use a stack to record the signals in the sensitivity lists in if/else statements and assign this set of signals to the registers within the same sequential block as the control signals. 
    Irrelevant signals in sensitivity lists that are not `controlling' (e.g.,  clock and reset are not control signals) are pruned.\\
    \noindent(II) \textit{Identify registers under improper control:}
    After pruning, if the control signal set is empty, the corresponding register is regarded as unprotected because untrusted users could modify the register by issuing write transactions directly. 
    We also compare the control signal sets for register arrays. 
    Since an array of signals should have the same `level' of protection, each entry should have the same number of control signals.
    Additionally, a warning will be raised if different entries do not have an identical set of control signals.

%% file: section/04Results.tex
\section{Experimental Work and Results}
\label{sec:results}

\subsection{Overview}
To evaluate the practicality and limitations of the \sol{} framework, we implement the scanners detailed in \autoref{sec:algorithms} and apply them on several open-source designs. 
Our experiments characterize several elements of the framework, including the number of instances caught by \sol{}, the wall-clock performance of the scanners, and a detailed study of the warnings produced for an \ac{SoC}. 
As part of our implementation, we curated a set of security-relevant keywords for signal names, the effect of which we also explore. 

\autoref{tbl:CWEAT-results-summary} lists the designs on which we evaluated \sol{}. 
The first four entries in the table are SoCs consisting of cores and peripherals and the remaining designs are processors only. 
The SoCs include several peripherals to show applicability of scanners on modules outside of the core. 
The designs have diversity in terms of complexity of the cores and HDL languages (Verilog and SystemVerilog) and cover two popular open-source ISAs (RISC-V and OpenRISC). 
As there is no ground truth about the presence of \acp{CWE}, we manually inspect the warnings in our detailed analysis (\autoref{sec:securityassesment}). 
We ran our experiments on an AMD Ryzen 9 3950X 16-Core Processor, 64~GB DDR4 RAM, using Ubuntu 20.04. The framework uses Verific libraries (academic license), gcc 9.4.0, and Python 3.8.10. 

\subsection{Initial Analysis}
\autoref{tbl:CWEAT-results-summary} summarizes the scanning results. 
The scanners analyzed as few as 1 to as many as 409 files per project, with up to $\sim$76k lines of code (LoC). 
The scanner for CWE-1245 found the most potential instances of that \ac{CWE} across the projects, while the scanner for CWE-1234 found the least. 
The mean running time for scanning was $\sim$63~ms; producing the \acp{AST} took around $3\times$ the time of running the collection of scanners on average, suggesting that these static analyzers could provide fast feedback to designers.  For the Hack@DAC'21 SoC, we produce 53 security warnings; this is more targeted than the 597 warnings from a commercial tool (\autoref{sec:background-and-related-work}).


\begin{table}[h!]
\footnotesize
    \caption{\sol{} Results Summary}
    \vspace{-2mm}
    \label{tbl:CWEAT-results-summary}
    \renewcommand{\arraystretch}{0.8}
    \resizebox{\linewidth}{!}{%
        \begin{tabular}{L{2cm}C{1.2cm}C{1.2cm}C{1.1cm}C{0.24cm}C{0.24cm}C{0.24cm}C{0.24cm}C{0.24cm}}
        \toprule
        Design &  Files/LoC(k) analyzed & Files/LoC(k) not analyz.  & Time (ms) Analysis/ Scan  & \rotatebox{90}{\text{1234}} & \rotatebox{90}{\text{1271}} & \rotatebox{90}{\text{1245}} & \rotatebox{90}{\text{1280}} & \rotatebox{90}{\text{1262}} \\ 
        \midrule
        H@DAC21 (Openpiton)~\cite{hackevent_hackdac21_2022,dessouky_hardfails_2019}  & 328 / 67.3 & 112 / 18.2 & 354 / 76 & 2 & 13 & 17 & 12 & 9 \\ \midrule
        H@DAC18 (Pulpissimo)~\cite{hackevent_hackdac21_2022,dessouky_hardfails_2019} & 409 / 76.3 & 116 / 16.2 & 322 / 142 & 0 & 9 & 26 & 20 & 10 \\ \midrule
        Hummingbirdv2 (E203)~\cite{nuclei_system_technology_hummingbirdv2_2022}  & 128 / 48.3 & 8   / 1.8 & 254 / 27 & 0 & 1 & 2 & 0 & 2 \\ \midrule
        Orpsoc~\cite{openrisc_community_orpsoc-cores_2021}  & 181 / 61.2 & 23  / 6.7  & 229 / 175 & 0 & 0 & 21 & 0 & 0 \\ \midrule
        EH1 Swerv~\cite{western_digital_eh1_2022}  & 1   / 15.9 & 0   / 0 & 744 / 142 & 0 & 0 & 0 & 0 & 0 \\ \midrule
        Ibex~\cite{noauthor_ibex_2022}  & 34  / 7.6 & 3   / 1.4 & 34 / 44 & 2 & 1 & 0 & 0 & 0 \\ \midrule
        CV32E40P~\cite{gautschi_near-threshold_2017}  & 109 / 25.4 & 15  / 4.1 & 111 / 21 & 5 & 4 & 5 & 0 & 0 \\ \midrule
        SCR1~\cite{noauthor_scr1_2022} & 41  / 1.4 & 4   / 0.3 & 120 / 14 & 0 & 12 & 3 & 0 & 0 \\ \midrule
        SERV~\cite{kindgren_serv_2022} & 15  / 2.1 & 0   / 0 & 8 / 1 & 0 & 0 & 0 & 0 & 0 \\ \midrule
        Ultraembedded~\cite{ultraembedded_risc-v_2022} & 17  / 5.7 & 1   / 0.5 & 28 / 7 & 0 & 0 & 0 & 0 & 0 \\ \midrule
        Mor1kx~\cite{noauthor_mor1kx_2022}  & 46  / 20.7 & 3   / 0.7 & 95 / 53 & 0 & 3 & 0 & 1 & 0 \\ 
        \bottomrule
        \end{tabular}
    }
\end{table}

To measure the portion of designs relevant to \acp{CWE}, we consider the number of `relevant' \ac{AST} nodes visited when running the different scanners. A node is considered `relevant' if it needs to be traversed to detect the \ac{CWE}. We also compare the number of relevant nodes visited by our scanners when keywords are used versus not used.
This is shown in \autoref{tbl:ast-searchspace}.
The number of nodes traversed with keyword matches are, on average, 80.4$\times$ fewer than the number of nodes in \acp{AST} and 16.3$\times$ fewer than nodes relevant to CWEs. 
Clearly, the keyword list used as part of scanning is an important aspect of making \sol{} practical. 

\begin{table}[t]
    \caption{Fraction of relevant nodes for scanners. (r) is the \% of nodes traversed for the scanner. (r+kw) is the \% of nodes traversed because of keyword matches. }
    \vspace{-2mm}
    \label{tbl:ast-searchspace}
    \resizebox{\linewidth}{!}{%
        \centering
        \begin{tabular}{l @{\extracolsep{3pt}}c c c c c c @{} c c c}
        \toprule
       \multirow{2}{3em}{Design} &  \multicolumn{2}{c}{1234} & \multicolumn{2}{c}{1262} & \multicolumn{2}{c}{1271} & 1245 & 1280 & Total\\
        \cline{2-3} \cline{4-5} \cline{6-7}
        &r (\%) & r+kw (\%) & r(\%) & r+kw(\%) & r(\%) & r+kw(\%) & r(\%) & r(\%) & nodes (k) \\
        \midrule
        H@21	&0.815	&0.001	&25.3	&6.2	&20.9	&0.89	&2.49	&39.5	&228	\\
        H@18	&1.113	&0	&25.6	&6.42	&23.6	&1.41	&2.6	&36.3	&236	\\
        e203	&0.361	&0	&22.1	&1.4	&25.8	&0.55	&0.84	&29.1	&182	\\
        orpsoc	&1.897	&0	&29.2	&8.79	&29.7	&2.27	&3.04	&36.5	&118	\\
        swerv	&0.235	&0	&0	&35.69	&0.6	&35.71	&0	&53.2	&1030	\\
        ibex	&0.744	&0.012	&26.4	&4.79	&25.8	&0.72	&1.44	&40.1	&41	\\
        cv32	&1.188	&0.018	&27.9	&8.4	&21.3	&0.77	&2.55	&42.4	&76	\\
        scr1	&0.67	&0	&19.8	&4.22	&19.5	&0.88	&1.44	&29.2	&60	\\
        serv	&0.646	&0	&26.1	&3.97	&21.9	&0.62	&0	&37.2	&6	\\
        ultraemb.	&1.912	&0	&26.7	&8.56	&21.8	&2.66	&1.18	&39.8	&15	\\
        mor1kx	&1.782	&0	&27.5	&6.74	&25.4	&1.63	&0.95	&37.7	&56	\\
        \bottomrule
       \end{tabular}
    }
\end{table}

\begin{figure}[h!]
\centering
\hspace*{-0.15in}
\includegraphics[width=1\columnwidth]{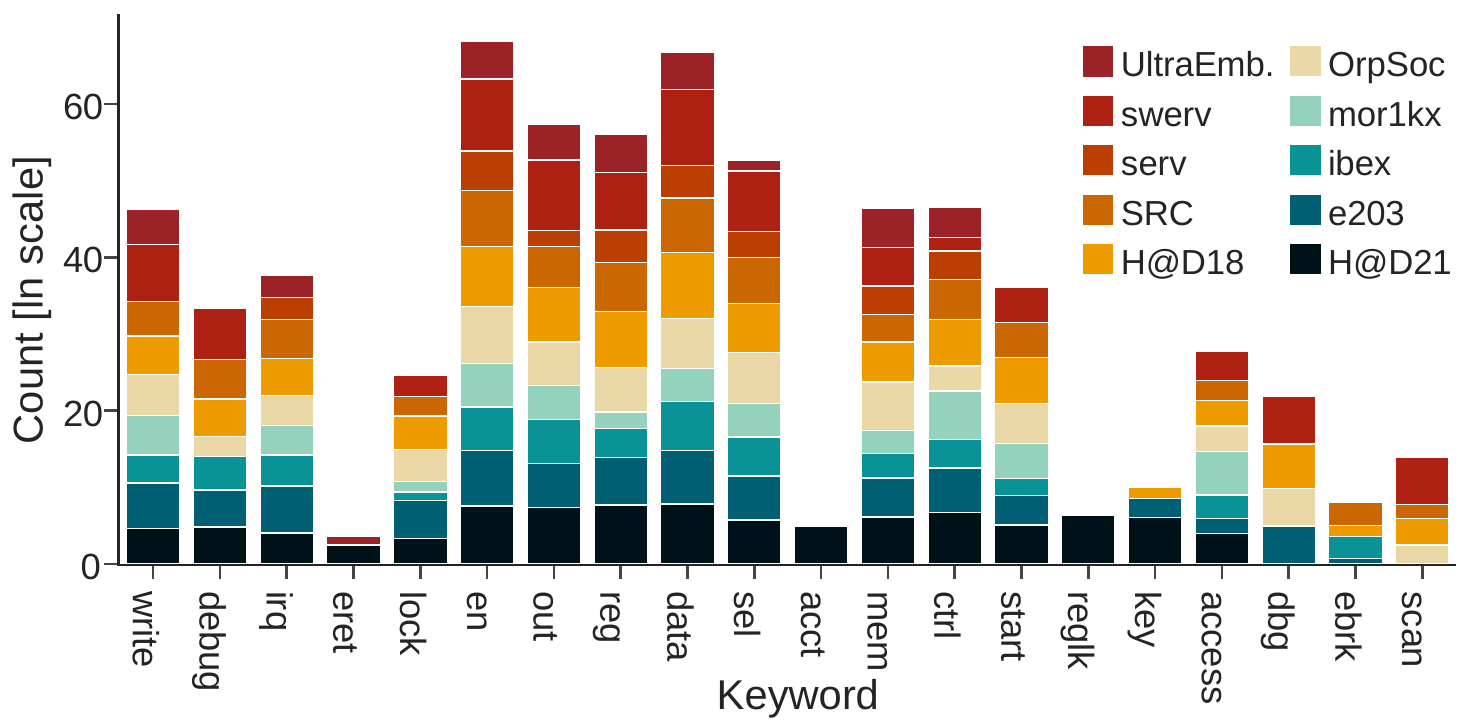}
\vspace{-9pt}
\caption{Keyword frequency in select open-source HW}
\label{fig:keywords}
\end{figure}

We devised our set of keyword-matching rules based on our experience with various design styles. This includes our initial experiments with open source designs, review of bug examples in literature \cite{bidmeshki_hunting_2021, dessouky_hardfails_2019, meng_rtl-contest_2021, tyagi_thehuzz_2022} and participation in hardware bug hunting competitions.
For additional insight, we analyzed how often signals match our keywords across the different open-source projects. 
The results are reported in \autoref{fig:keywords}. 
The frequency values are reported in log scale due to the high range of total identifiers across designs.
The keywords cover a good portion of the considered benchmarks, indicating broad applicability for a framework like \sol{} to be used out-of-the-box in different projects. 
While our keywords are meaningful across the designs that we considered, our goal is not to provide the best set of keywords, but to show that keywords are an effective way to build security scanners. 
In real-world applications, performance can be improved by designers modifying keywords. 

%% file: section/05SecurityAssesment.tex
\subsection{Detailed Analysis}
\label{sec:securityassesment}


\label{subsubsec:findings-classification}
We evaluate each warning picked up by \sol{} and classify it as True Positive (\textbf{T}), Indeterminate (\textbf{I}), or False Positive (\textbf{F}). 
\textbf{T} means we confirmed 
the \ac{CWE}'s presence. 
Ascertaining if a \ac{CWE} leads to an exploitable vulnerability is out of scope of this analysis. 
Conversely, \textbf{F} means 
we determined that the \ac{CWE} was incorrectly flagged. 
In cases where we could not discern an instance into \textbf{T} or \textbf{F}, we classified it as \textbf{I}. 
\autoref{fig:cwe_classification} shows our classification.



To explain our analysis and classifications of \sol{}'s warnings, 
we discuss examples found in the Hack@DAC-2021 SoC, demonstrating the use of \sol{} and providing insights into the practicalities of using this approach.
There are deliberate vulnerabilities implanted in this SoC, but we do not have access to a comprehensive list of those vulnerabilities. It is also not our goal to detect those deliberately inserted vulnerabilities.

\Com{
\begin{figure}
\centering
\begin{subfigure}{0.27\linewidth}
    \includegraphics[width=\linewidth]{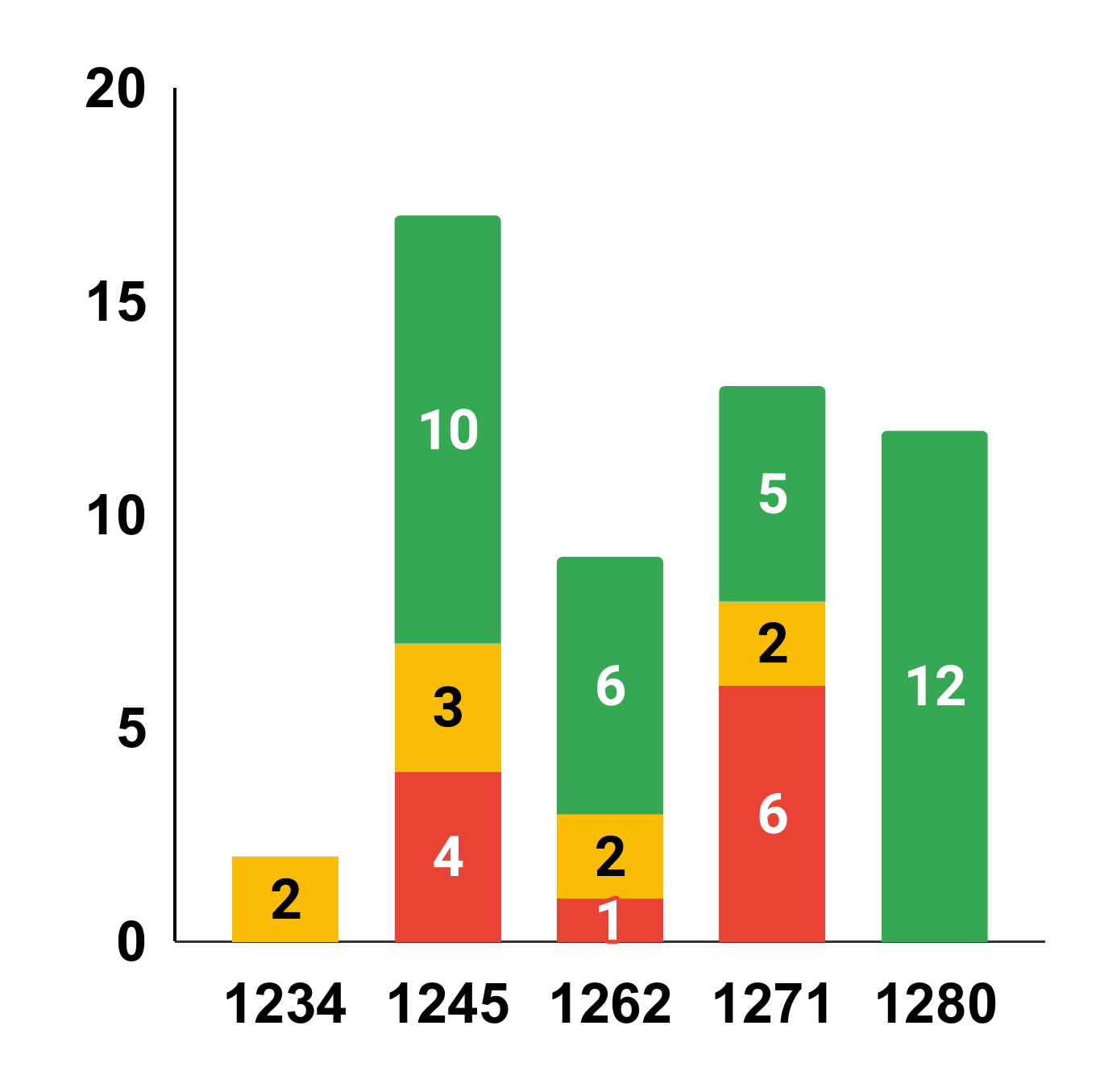}
    \caption{}
    \label{fig:h-21_classification}
\end{subfigure}
\hspace{-0.15in}
\begin{subfigure}{0.27\linewidth}
    \includegraphics[width=\linewidth]{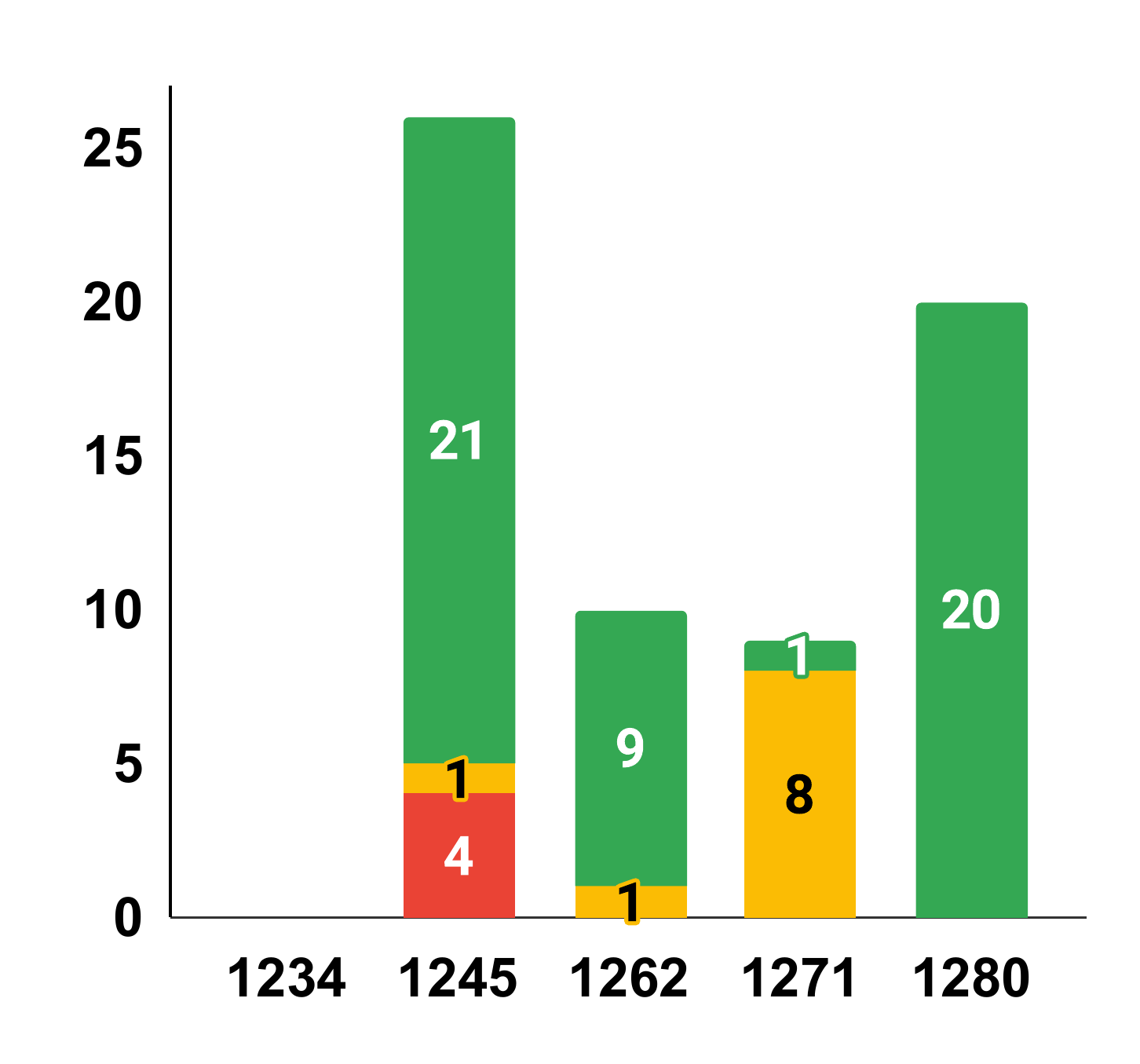}
    \caption{}
    \label{fig:h-18_classification}
\end{subfigure}
\hspace{-0.15in}
\begin{subfigure}{0.27\linewidth}
    \includegraphics[width=\linewidth]{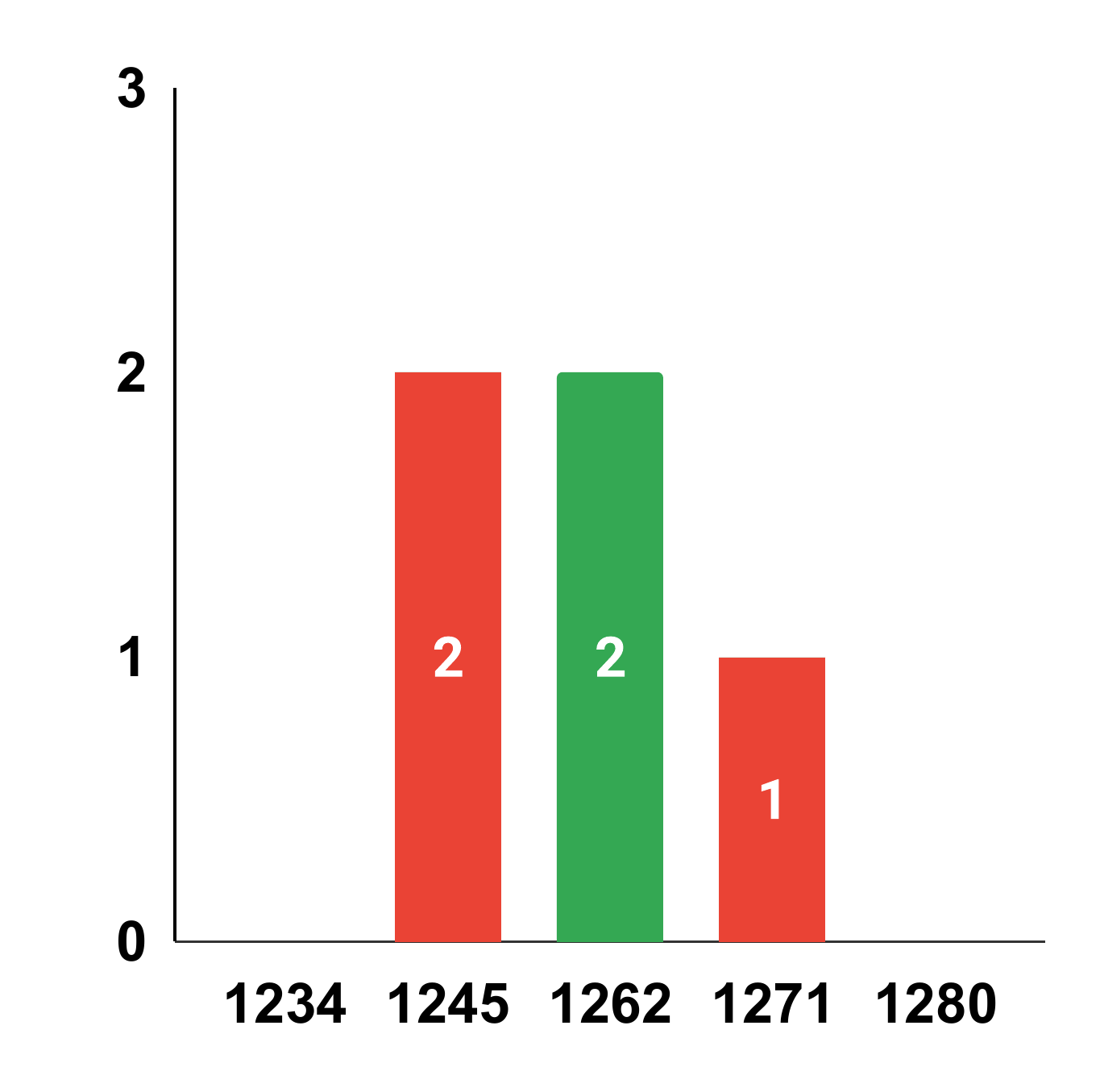}
    \caption{}
    \label{fig:e203_classification}
\end{subfigure}
\hspace{-0.15in}
\begin{subfigure}{0.27\linewidth}
    \includegraphics[width=\linewidth]{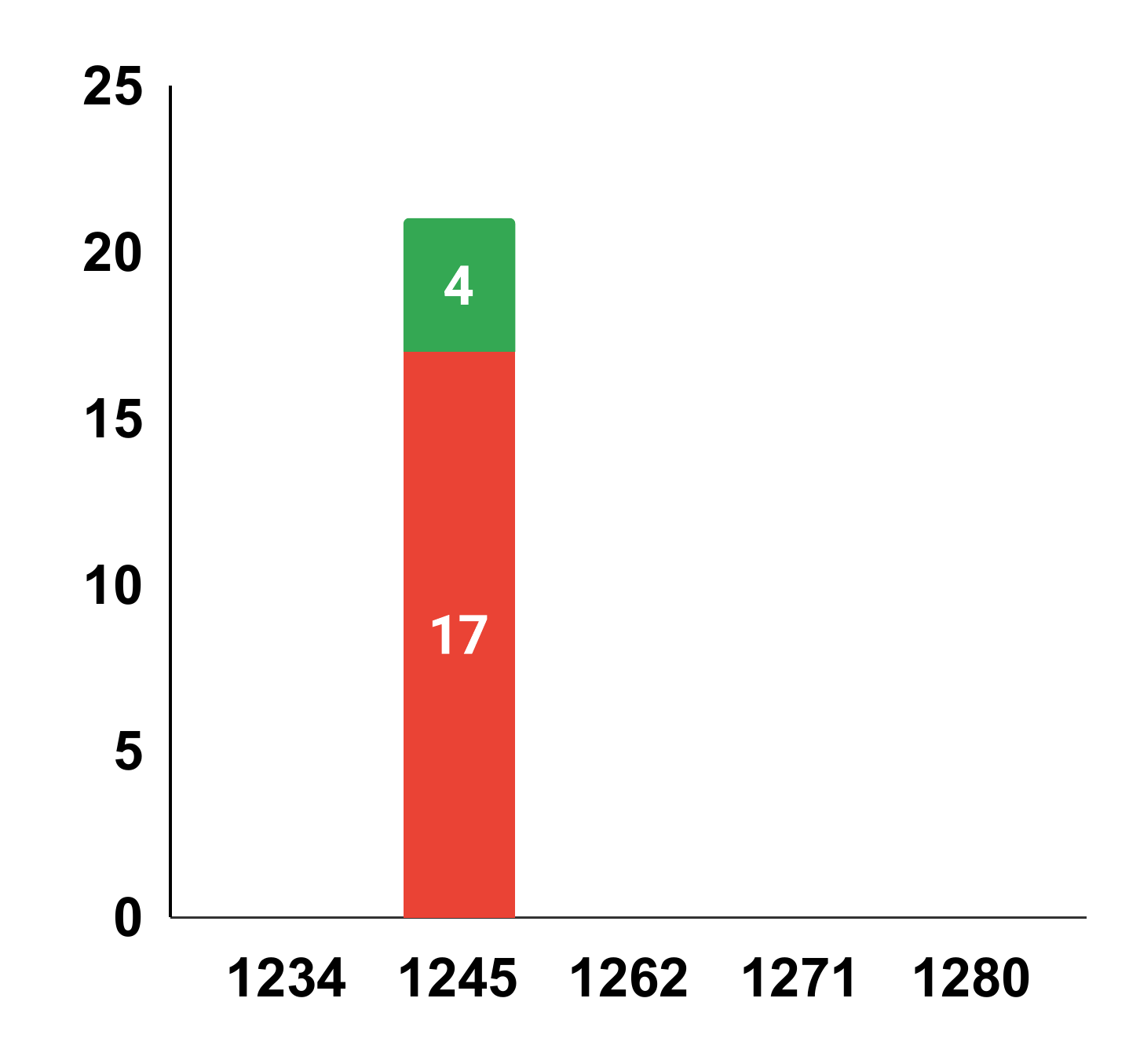}
    \caption{}
    \label{fig:orpsoc_classification}
\end{subfigure}
\vspace*{-0.1in}
\caption{CWE classification of instances caught by \sol{} in SoCs. (a) Hack@DAC2021 (b) Hack@DAC2018 (c)E203 (d) OrpSoC. Legend: Red (P-CWE Present), Yellow(I-Indeterminate), Green(N-CWE Not Present)}
\label{fig:cwe_classification}
\end{figure}
}

\Com{}
\begin{figure}[h!]
    \centering
    \includegraphics[width=\linewidth]{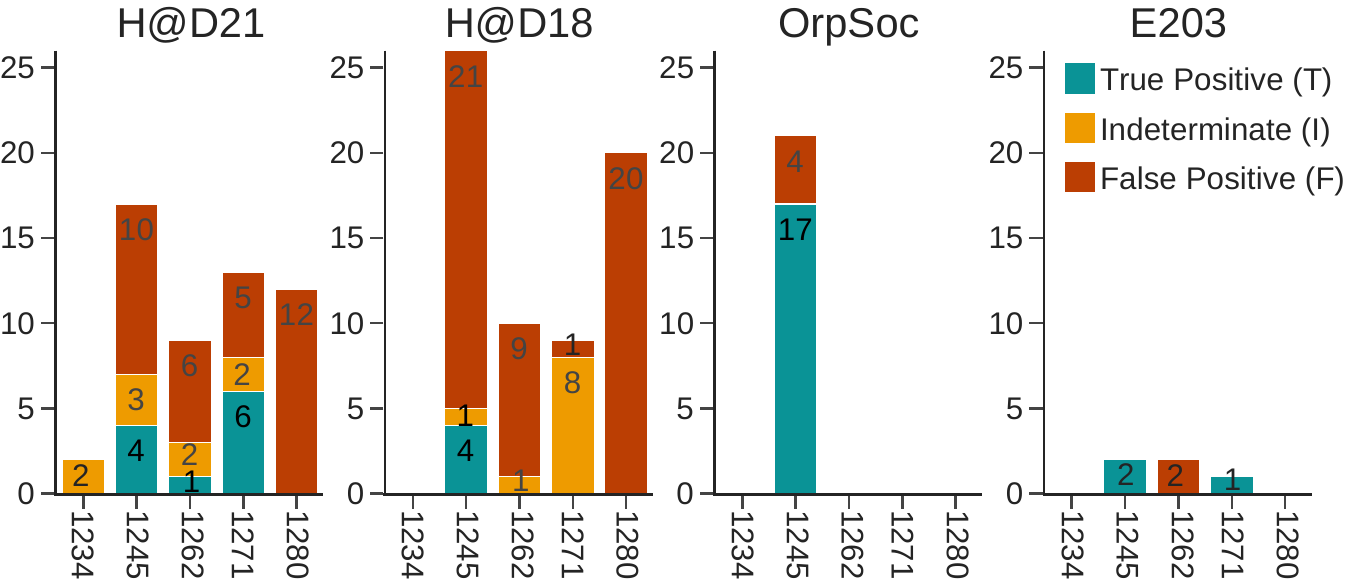}
    \vspace*{-0.2in}
    \caption{Classification of instances caught by \sol{} in SoCs.}
    \label{fig:cwe_classification}
\end{figure}

For \ac{CWE}-1234, two instances were identified (\autoref{fig:lst:1234}). 
\autoref{fig:lst:1234-instance1} is part of the flush controller for the processor. 
If the highlighted condition is met, the Instruction Fetch, Decode and Execution stages are flushed. 
This is understandable when an exception is raised or when returning from an exception. 
Flushing the pipeline is also expected while entering the debug mode, however, whether the debug request should override exception-related signals depends on the design specification. 
If the intended flush mechanism is different, this instance could represent a vulnerability. 
Similar analysis follows for \ref{fig:lst:1234-instance2} which pertains to the implementation of un-stalling the core in the cases of interrupt and debug.
We need more information for judging, hence we classify both warnings for the code in \autoref{fig:lst:1234} as \textbf{I}.


\begin{figure}[h]
\centering

    \begin{subfigure}[b]{\linewidth}
    \begin{lstlisting}[language=verilog, linebackgroundcolor={\ifnum\value{lstnumber}>0
                \ifnum\value{lstnumber}<2
                    \color{yellow}
                \fi
            \fi}]
    if (ex_valid_i || eret_i || set_debug_pc_i) begin
        set_pc_commit_o        = 1'b0;
        flush_if_o             = 1'b1;
        flush_unissued_instr_o = 1'b1;
        flush_id_o             = 1'b1;
        flush_ex_o             = 1'b1;
    \end{lstlisting}
    \vspace{-3mm}
    \caption{Debug overrides exception to flush pipeline in Controller module}
    \label{fig:lst:1234-instance1}
    \end{subfigure}
    
    \vspace{1mm}
    
    \begin{subfigure}[b]{\linewidth}
    \begin{lstlisting}[language=verilog, linebackgroundcolor={\ifnum\value{lstnumber}>2
                \ifnum\value{lstnumber}<4
                    \color{yellow}
                \fi
            \fi}]
    always_comb begin : wfi_ctrl // wait for interrupt register
        wfi_d = wfi_q;//if there is any interrupt pending un-stall the core also un-stall to enter debug mode
        if (|mip_q || debug_req_i || irq_i[1]) begin
            wfi_d = 1'b0;
        // or alternatively if there is no exception pending and we are not in debug mode wait here for the interrupt
        end else if (!debug_mode_q && csr_op_i == WFI && !ex_i.valid) begin
            wfi_d = 1'b1;
    \end{lstlisting}
    \vspace{-3mm}
    \caption{Debug overrides interrupt to un-stall core in Csr\_regfile module}
    \label{fig:lst:1234-instance2}
    \end{subfigure}

\vspace{-3mm}
\caption{Instances found by Scanner 1234. Highlighted are line(s) reported by scanner as a potential weaknesses}
\label{fig:lst:1234}
\vspace{-3mm}
\end{figure}

For \ac{CWE}-1271, \sol{} picks up 13 instances: 6 \textbf{T}, 2 \textbf{I} and 5 \textbf{F}. 
An example is shown in \autoref{fig:lst:1271} in which we classify registers \texttt{pmp\_access\_type\_en} and \texttt{pmp\_access\_type\_new} as \textbf{T} and \textbf{I} respectively. 
This snippet belonging to the Direct Memory Access (DMA) module, shows the use of registers containing sensitive information for Physical Memory Protection (PMP) regions. 
The PMP access registers set the permissions to read and/or write for regions in the memory. 
To change the value of this register, \texttt{pmp\_access\_type\_en} is used. 
It is a critical register but is not appropriately reset under a reset block. 
When the circuit is first brought out of reset, the state of \texttt{pmp\_access\_type\_en}  will be unknown. 
An attacker can deliberately bring the system in and out of reset to exploit this situation, hence we classify it as \textbf{T}. 
\texttt{pmp\_access\_type\_new} also does not have a proper reset on initialization. 
But since its sole purpose is to store the next value of \texttt{pmp\_access\_type\_reg}, which is appropriately reset, it cannot be clearly stated whether a weakness exists -- hence, \textbf{I}.

\begin{figure}[t]
\centering

    \begin{lstlisting}[language=verilog, linebackgroundcolor={\ifnum\value{lstnumber}>0
                \ifnum\value{lstnumber}<4
                    \color{yellow}
                \fi
            \fi}]
    riscv::pmp_access_t 
    pmp_access_type_reg, pmp_access_type_new; // riscv::ACCESS_WRITE or riscv::ACCESS_READ
    reg pmp_access_type_en;
    // code skipped for brevity
        if (pmp_access_type_en)
            pmp_access_type_reg <= pmp_access_type_new;

    \end{lstlisting}
    \vspace{-3mm}
  
\caption{Instances found by Scanner 1271 in DMA module.}
\label{fig:lst:1271}
\vspace{-3mm}
\end{figure}

For \ac{CWE}-1245, 17 instances were picked up by CWEAT: 4 \textbf{T}, 3 \textbf{I} and 10 \textbf{F}. 
Two \textbf{T} instances are shown in \autoref{fig:lst:1245-vul}. 
The first instance (\autoref{fig:lst:1245-vul-instance1}) concerns a deadlock for the modular exponentiation module used in RSA implementation. 
There is no transition out of the state \texttt{HOLD}. 
On inspection, we see that the transition out of \texttt{HOLD} is possible only on the reset signal received from the RSA top module. 
This reset signal, however, is a combination of a reset signal and an internal signal which is modified in the write process of the RSA peripheral. 
This may allow room for communication with the RSA peripheral to put \texttt{mod\_exp} in the \texttt{HOLD} state at will, causing denial of service. 
The second instance (\autoref{fig:lst:1245-vul-instance2}) has an incomplete case statement with \texttt{s13} not covered in the \ac{FSM} for an AES interface. 
This leaves room for an adversary to bring the system to a state for which functionality is undefined, resulting in denial of service. 
Hence, both these instances are classified as \textbf{T}.
However, we  classified many instances picked up by \sol{} as \textbf{F}. 
This happens because particular styles of \ac{FSM} implementation are not yet covered by 1245 scanner (e.g., multiple signals used in case condition, hard-coded one-hot encoded state variables or use of ternary operators). 
Instances were classified as \textbf{I} if they were not false positives and either had limited context or were too complex.

\begin{figure}[h]
\centering

    \begin{subfigure}[b]{\linewidth}
    \begin{lstlisting}[language=verilog, linebackgroundcolor={\ifnum\value{lstnumber}>5
                \ifnum\value{lstnumber}<8
                    \color{yellow}
                \fi
            \fi}]
    `UPDATE: begin
        if(exponent_reg!='d0) begin //code skipped for brevity
        end
        else state <= `HOLD;
    end
    `HOLD: begin
        end
    \end{lstlisting}
    \vspace{-3mm}
    \caption{\ac{FSM} deadlock at state HOLD in mod\_exp module}
    \label{fig:lst:1245-vul-instance1}
    \end{subfigure}
    
    \vspace{-1mm}
    
    \begin{subfigure}[b]{\linewidth}
    \begin{lstlisting}[language=verilog, linebackgroundcolor={\ifnum\value{lstnumber}>3
                \ifnum\value{lstnumber}<5
                    \color{yellow}
                \fi
            \fi}]
        s12: begin
           dii_data_vld  <= 1'b0;
           state <= s14;
        end
        s14: begin
           if(ct_valid == 1) state <= s15;
           else state <= s14;
        end // code skipped for brevity
    \end{lstlisting}
    \vspace{-3mm}
    \caption{Incomplete case statement in aes2\_interface module}
    \label{fig:lst:1245-vul-instance2}
    \end{subfigure}

\vspace{-3mm}
\caption{Instances found by Scanner 1245.}
\label{fig:lst:1245-vul}
\vspace{-3mm}
\end{figure}

For \ac{CWE}-1280, all detected instances were verified to be false positives (\textbf{F}). \autoref{fig:1280example} shows an example,  
taken from the modulo operator of the RSA module of the Ariane core. It highlights how \texttt{p} is accessed before being updated, but in this instance, it follows the expected implementation and is not performing an access on an asset.
Simple instances of \ac{CWE}-1280 suggest flaws in the design and are unlikely to pass verification, while complex occurrences require more context to be detected. 

\begin{figure}[h]
    \centering
    \begin{lstlisting}[language=verilog,linebackgroundcolor={\ifnum\value{lstnumber}>5
                \ifnum\value{lstnumber}<7
                    \color{yellow}
                \fi
            \fi 
            \ifnum\value{lstnumber}>3
                \ifnum\value{lstnumber}<5
                    \color{yellow}
                \fi
            \fi}]
        p = {p[WIDTH-2:0],A[WIDTH-1]};
        A[WIDTH-1:1] = A[WIDTH-2:0];
        p = p-N;
        if(p[WIDTH-1] == 1)begin
            A[0] = 1'b0;
            p = p + N;   
    \end{lstlisting}
    \vspace{-3mm}
    \caption{False positive detected by Scanner 1280.}
    \label{fig:1280example}
\end{figure}

For \ac{CWE}-1262, \sol{} picks up 9 instances: 1 \textbf{T}, 2 \textbf{I} and 6 \textbf{F}. 
One \textbf{T} instance is shown in~\autoref{fig:reglk_code}.
As specified in \autoref{sec:1262}, we first collect a set of control signals for \texttt{reglk\_mem[0]}, which contains \texttt{jtag\_unlock, en, we, reglk\_ctrl[3]}. 
Note that the signals associated with \texttt{clk} and \texttt{reset} are pruned out.
Since there are still four control signals for \texttt{reglk\_mem[0]}, it is write-protected. 
However, when we compare the control signals for the \texttt{reglk\_mem} array, only \texttt{reglk\_mem[0]} is controlled by \texttt{reglk\_ctrl[3]}, while other entries are controlled by \texttt{reglk\_ctrl[1]}, as highlighted in \autoref{fig:reglk_code}.
This difference might not be an issue if designers made it intentionally.
Designers might use different signals to control a signal array, as shown in \autoref{fig:acct_code} where each entry of \texttt{acct\_mem} is controlled by different \texttt{reglk\_ctrl} values.
Since each entry of \texttt{acct\_mem} represents different IPs' access control values, and each \ac{IP} has different security levels, this array may not have the same control signals.
However, in \autoref{fig:reglk_code}, only one entry of \texttt{reglk\_mem} has different control signals, without evidence to justify its legality and is classified as CWE-1262. We present an attack scenario
to exploit this bug. 

\subsection{User Exploit for CWE-1262}
\label{sec:user_exploit}
When we scanned the Hack@DAC-2021 SoC, they pointed to some files to potentially contain CWEs, including the one shown in \autoref{fig:reglk_code} where CWE-1262 is present.
To verify the validity, we generate an attack scenario to confirm the vulnerability.
As the code snippet shows, \texttt{reglk\_mem[0]} is controlled by \texttt{reglk\_ctrl[3]}, while other bits of \texttt{reglk\_mem} are controlled by \texttt{reglk\_ctrl[1]}.
This difference could be used by attackers if designers are not aware of it and set the \texttt{reglk\_ctrl[3]} and \texttt{reglk\_ctrl[1]} to 0 and 1, respectively.
The \texttt{reglk\_mem} stores the control values for the registers in peripheral IPs, which appears to be \texttt{reglk\_ctrl} in the modules.
Once the protection for \texttt{reglk\_mem} is bypassed, these control values can be reconfigured and several IPs' registers can be modified by attackers, resulting in the security issue.

We set the \texttt{REGLK\_REGLK} value in \texttt{reglk.h} to \texttt{0xf6}, which makes \texttt{reglk\_ctrl[3]} 1 and allows untrusted users to override \texttt{reglk\_mem[0]} at run-time.
We ran the simulation on Hack@DAC-2021 platform and issued a write transaction to bypass \texttt{reglk\_ctrl} protection to modify \texttt{reglk\_mem[0]}, leaving security-relevant signals in AES0, AES1, and SHA256 prone to leaks and modifications.

\begin{figure}[t]
    \centering
    \begin{subfigure}[a]{\linewidth}
    \begin{lstlisting}[language=verilog,linebackgroundcolor={\ifnum\value{lstnumber}=3
                    \color{yellow}
            \fi}]
else if(en && we)
    case(address[7:3])
    0: reglk_mem[0]  <= reglk_ctrl[3] ? reglk_mem[0] : wdata;
    1: reglk_mem[1]  <= reglk_ctrl[1] ? reglk_mem[1] : wdata; 
    // code skipped for brevity
    5: reglk_mem[5]  <= reglk_ctrl[1] ? reglk_mem[5] : wdata;
    default:;

    \end{lstlisting}
    \vspace{-3.5mm}
    \caption{A \textbf{T} instance found in reglk\_ctrl wrapper module.}
    \label{fig:reglk_code}
    \end{subfigure}
    \begin{subfigure}[b]{\linewidth}
    \begin{lstlisting}[language=verilog,linebackgroundcolor={
            \ifnum\value{lstnumber}=2
                    \color{yellow}
            \fi
            \ifnum\value{lstnumber}=4
                    \color{yellow}
            \fi
            \ifnum\value{lstnumber}=6
                    \color{yellow}
            \fi
            \ifnum\value{lstnumber}=8
                    \color{yellow}
            \fi}
            ]
case(address[10:3])
0: acct_mem[00]  <= reglk_ctrl[5] ? acct_mem[00] : wdata;
1: //code skipped for brevity
3: acct_mem[03]  <= reglk_ctrl[13] ? acct_mem[03] : wdata;
4: //code skipped for brevity
6: acct_mem[06]  <= reglk_ctrl[1] ? acct_mem[06] : wdata;
7: //code skipped for brevity
9: acct_mem[09]  <= reglk_ctrl[7] ? acct_mem[09] : wdata;
default:;

    \end{lstlisting}
    \vspace{-3mm}
    \caption{An \textbf{I} instance found in acct\_ctrl wrapper module.}
    \label{fig:acct_code}
    \end{subfigure}
    \vspace{-6mm}

    \caption{Instances found by Scanner 1262.}
    \vspace{-6mm}

\end{figure}

%% file: section/06Discussion.tex
\section{Discussion and Limitations\label{sec:dis}}

While developing our scanners, we faced a challenge: balancing accuracy (ratio of true positives over total instances caught by scanner) and comprehensiveness (number of true positives identified). 
This trade-off exists because, to increase comprehensiveness, we need to cast a wider net (increase keyword breadth) and traverse more source code. 
This will identify more weaknesses but can also flag more false positives. 
A decrease in accuracy of the tool makes it less practical to use. 
Conversely, if the goal is to more accurately identify weaknesses, some true positives will not be picked up, increasing the probability of weaknesses persisting in designs. 
While this trade-off is an important consideration for static analysis tools, the goal of this work is not to provide the best scanners, but to show that such \textbf{static scanners can be implemented in a way that can be useful for designers to write more secure modules}. 

An approach to improve the scanners could involve ``learning'' the coding style of a given project, i.e., using user feedback to update the scanning. 
Interaction between designers and warnings generated can be used to iteratively increase accuracy. 
However, caution must be taken to avoid the pitfall of designers providing poor feedback (e.g., marking everything safe to avoid alarm fatigue). 
An example user-driven refinement is shown in \autoref{ tbl:evolution_det_alg}, which shows the results after revising keywords in the scan of the Hack@DAC 2021 SoC. 
By default, we scan the design using a default list (first row). 
This includes matches with sensitive signal names in the \texttt{true} list and prunes out matched with the \texttt{false} list. 
The user can add signals they consider `security-relevant'(second row). 
This results in more hits. The user can then mark certain signals as irrelevant by appending to the \texttt{false} list (third and fourth row).
In this case, one true weakness is missed, but drastically improves the number of true positives as a fraction of the total number of instances/hits caught by the scanner.

\begin{table}[t]
    \caption{Evolution of 1271 scanner with user feedback}
    \vspace{-2mm}
    \label{ tbl:evolution_det_alg}
    \resizebox{\linewidth}{!}{%
        \centering
        \begin{tabular}{l l  c  c  c  c  c}
        \toprule
        User Interaction & Keywords & \# hits & \textbf{T} & \textbf{I} & \textbf{F} & \textbf{T}/ \# hits\\
        \midrule
        None & true:\textcolor{blue}{$\{lock,prot\}$} false:\textcolor{red}{$\{clock\}$}&66&2&17&47&0.031\\
        \midrule
        Add `access' & true:\textcolor{blue}{$\{lock,prot,access\}$} false:\textcolor{red}{$\{clock\}$ } &75&7&19&49&0.095\\
        \midrule
        Prune `block'& true:\textcolor{blue}{$\{lock,prot,access\}$} false:\textcolor{red}{ $\{clock,block\}$}&47&6&11&30&0.128\\
        \midrule
        \multirow{2}{6em}{Prune `ar\_lock' `aw\_lock' } & \multirow{2}{16em}{true:\textcolor{blue}{$\{lock,prot,access\}$} false:\textcolor{red}{$\{clock,block,ar,aw\}$} }&18&6&2&10&0.333\\
        &&&&\\
        \bottomrule
      \end{tabular}
    }
\end{table}

\Com{
\todoblock{
\begin{enumerate}
    \item some contexts may be needed to reduce the false positive rates
    \item limitation of the current algorithms
    \item learn users' habits/coding style as keywords
    \item Simple occurrences of \ac{CWE} 1280 require major flaws in the
design and are unlikely to pass verification steps, while more
complex occurrences are hard to detect.
\end{enumerate}
}}

Our approach cannot guarantee the absence of a weakness when no warnings are produced. 
These `false negatives' exist because we use heuristic detection of \textit{patterns}. 
In our experiments, there are no benchmarks for comparison as we are in the initial stages of classifying, gathering, and understanding weaknesses in RTL. 
Our experimental evaluation focused on selected open-source designs with subsequent manual inspection to gauge effectiveness of scanners. 
%
%
%
%
%
%
%
In our manual inspection, we are somewhat limited when evaluating others' designs without complete context to understand intended implementation. 
Since documentation is limited and source code comments reveal partial information, evaluating instances picked up by scanners is limited by the expertise of the security evaluator. 
That said, if the examiner is also the designer, classification should be easier and faster. 
In an industry setting, there will be fewer uncertainties as documentation that conveys security-relevant information is more readily available. Design and security teams can work together to navigate potential concerns. 



%% file: section/07Conclusion.tex
\section{Conclusions and Future Work\label{sec:conc}\label{sec:conclusions}}

We have shown that certain \acp{CWE} can be detected during the early RTL Implementation stage. In the 11 open-source designs, 180 instances were caught by \sol{}. We evaluated the 144 found in SoCs and confirmed weakness in 35 (24.31\%) of the instances. 19 (13.1\%) were indeterminate and 90 (62.5\%) were not weaknesses.
We have also shown that CWEAT restricts the search for particular weaknesses to a smaller subset. Instead of having to go over thousands of lines of source code, the search for specific issues is limited to a few signals and modules. By adding new scanners, other \acp{CWE} could be analyzed, increasing the level of security of the design.

The \acp{CWE} are varied in nature and can be detected at various stages. This provides two concrete directions for research. Firstly, scanners can be developed for the 21 \acp{CWE} that can be detected statically with specification files as shown in Table \ref{tbl:classification-CWEs}. Secondly, more accurate and comprehensive scanners can be developed for the \acp{CWE} we have targeted. Both  approaches aim to standardize detection of hardware weaknesses, which we believe is significant.



\begin{acks}
We thank Verific Design Automation for generously providing academic access to linkable libraries, examples, and documentation for their RTL parsers.
This research work is supported in part by a gift from Intel Corporation. This work does not in any way constitute an Intel endorsement of a product or supplier.
We acknowledge the support of the Natural Sciences and Engineering Research Council of Canada (NSERC), RGPIN-2022-03027. 

\end{acks}
